\documentclass[12pt]{article}
\usepackage{times}
\usepackage{enumerate}
\usepackage{comment}
\usepackage{url}
\usepackage{subfig}
\usepackage{amsmath}
\usepackage{amsthm}
\usepackage{amssymb}
\usepackage{wrapfig}
\usepackage{graphicx}
\graphicspath{{./Figs/}}
\usepackage{paralist}
\usepackage[active]{srcltx}
\usepackage{algorithm}
\usepackage{fullpage}

\newenvironment{proof sketch}[1]{\noindent {\emph{Proof sketch of #1:}}}{\hfill \qed}
\newtheorem{theorem}{Theorem}

\newtheorem{lemma}{Lemma}
\newtheorem{definition}{Definition}

\newtheorem{rem}{Remark}
\newtheorem{observation}{Observation}

\newcommand{\eqdf}{\triangleq}
\newcommand{\eps}{\varepsilon}

\newcommand{\res}{m}

\newcommand{\cmax}{c_{\max}}
\newcommand{\cmin}{c_{\min}}
\newcommand{\bmax}{b_{\max}}
\newcommand{\dmax}{d_{\max}}

\newcommand{\M}{\max\{\lambda,\mu\}}

\newcommand{\bpb}{\textit{bpb}}

\newcommand{\MCF}{\textsc{mcf}}
\newcommand{\ONMCF}{\textsc{onmcf}}
\newcommand{\dlp}{\textsc{d-lp}}
\newcommand{\plp}{\textsc{p-lp}}
\newcommand{\val}{\emph{\text{benefit}}}
\newcommand{\valp}{\emph{\text{value}}}
\newcommand{\I}{F^*}

\newcommand{\cost}{\emph{\text{cost}}}

\newcommand{\alg}{\textsc{alg}}
\newcommand{\oracle}{\emph{\text{oracle}}}
\newcommand{\opt}{\textsc{min-cost}}

\newcommand{\PR}{\Delta_j P}
\newcommand{\DU}{\Delta_j F}

\begin{document}

\title{Online Multi-Commodity Flow with High Demands}

\author{%
Guy Even\thanks{School of Electrical Engineering, Tel-Aviv
Univ., Tel-Aviv 69978, Israel.
\protect\url{{guy,medinamo}@eng.tau.ac.il}} \and Moti Medina$^*$
}
%\date{}
\maketitle

\begin{abstract}
  This paper deals with the problem of computing, in an online
  fashion, a maximum benefit multi-commodity flow (\ONMCF), where the
  flow demands may be bigger than the edge capacities of the network.

  We present an online, deterministic, centralized, all-or-nothing, bi-criteria algorithm. The competitive ratio of the
  algorithm is constant, and the algorithm augments the capacities by at
  most a logarithmic factor.

  The algorithm can handle two types of flow requests: (i) low demand
  requests that must be routed along a path, and (ii) high demand
  requests that may be routed using a multi-path flow.

  Two extensions are discussed: requests with known durations and
  machine scheduling.
\end{abstract}

\paragraph{Keywords.}
Online algorithms, primal-dual scheme, multi-commodity flow.

\section{Introduction}
We study the problem of computing a multi-commodity flow in an online
setting (\ONMCF).  The network is fixed and consists of $n$ nodes and
$\res$ directed edges with capacities.
The adversary introduces flow requests in an online fashion.

A flow request $r_j$ is specified by the source node $s_j$, the target
node $t_j$, the demand $d_j$, i.e., the amount of flow that is
required, and the benefit $b_j$, i.e., the credit that is given for a
served request.

We focus on an \emph{all-or-nothing} scenario, where a credit $b_j$ is
given only if a request $r_j$ is fully served, otherwise, the credit
is zero.  Given a sequence of flow requests, the goal is to compute a
multi-commodity flow (\MCF) that maximizes the total benefit of
\emph{fully served} requests. Our algorithm can deal with high
demands $d_j$.  In particular, the demand $d_j$ may be
bigger than the maximum capacity.

\paragraph{Our Contribution.}
We present a centralized, deterministic, all-or-nothing,
non-preemptive online algorithm for the \ONMCF\ problem with high
demands.  The algorithm is $O(1)$-competitive. The algorithm violates
edges capacities by an $O(\log n)$ factor.

We show how to extend the algorithm so that it handles two types of flow requests: (i) low demand requests that must be routed along a path, and (ii) high demand requests that may be routed using a multi-path flow.

Finally, two extensions are discussed: requests with known durations and machine scheduling.

\subsection{Previous Work}
Online multi-commodity flow was mostly studied in the context of \emph{single path routing}.
The \emph{load} of an edge $e$ in a network is the ratio between the flow that traverses $e$ and its capacity.

Online routing was studied in two settings: (1)~throughput
maximization, i.e., maximizing the total benefit gained by flow
requests that are served~\cite{AAP,BN06,DBLP:conf/icdcn/EvenMSS12},
and (2)~load minimization, i.e., routing all requests while minimizing
the maximum load of the
edges~\cite{aspnes1997line,awerbuch2001competitive,BN06,DBLP:conf/podc/BansalLNZ11}.

In these two settings the following variants are considered: (1)~permanent routing~\cite{aspnes1997line,BN06,DBLP:conf/icdcn/EvenMSS12}, (2)~ unknown durations~\cite{awerbuch2001competitive}, and (3)~known durations~\cite{AAP,DBLP:conf/icdcn/EvenMSS12,DBLP:conf/podc/BansalLNZ11}.

\paragraph{Load Minimization.}
In the case of permanent routing, Aspnes et. al~\cite{aspnes1997line}
designed an algorithm that augments the edge capacities by a factor of
at most $O(\log n)$ w.r.t. a feasible optimal routing.  Buchbinder and
Naor~\cite{BN06} obtained the same result by applying a primal-dual
scheme.  This result can be extended to requests with high demands.
The extension is based on a min-cost flow oracle that replaces
the shortest path oracle.

Aspnes et. al~\cite{aspnes1997line} also showed how to use
approximated oracles to allocate Steiner trees in the context of
multicast virtual circuit routing. They obtained a competitive ratio
of $O(\log n)$.  Recently Bansal
et.al~\cite{DBLP:conf/podc/BansalLNZ11} extended this result to
bi-criteria oracles and showed how to embed $d$-depth trees and
cliques in the context of resource allocation in cloud computing. In
the case of cliques, they required that the pairwise demands are
uniform and smaller than the edge capacities.  For the case of clique
embedding they obtained a competitive ratio of $O(\log^3 n \cdot \log
(nT))$ w.r.t. a feasible optimal solution, where $T$ is the ratio
between the maximum duration to the minimum duration of a request.

\paragraph{Throughput Maximization.}
For the case of known durations, Awerbuch et. al~\cite{AAP} designed an
$O(\log (nT))$ competitive algorithm, where $T$ is the maximum request
duration.  This algorithm requires that the demands are smaller than
the edge capacity by a logarithmic factor.  Buchbinder and
Naor~\cite{BN06,BN09} introduced the primal-dual scheme in the online
setting and designed a bi-criteria algorithm that is $1$-competitive
while augmenting edge capacities by a factor of $O(\log n)$ for the
case of unit demands and unit benefits.

Recently Even et. al~\cite{DBLP:conf/icdcn/EvenMSS12} showed how to
apply the primal-dual scheme to embed a variety of traffic patterns in
the context of Virtual Networks (VNETs). Their goal is to maximize the
profit of the served VNET requests. Some of the results
in~\cite{DBLP:conf/icdcn/EvenMSS12} require solving the \ONMCF\
problem with high demands.
\begin{comment}
  Moreover, they showed in this paper how to reduce an algorithm to
  permanent routing to known duration.(verify that it works for
  tri-criteria!)
\end{comment}

\begin{comment}
``A second extension applies to the case where request $r_i$ has an arbitrary demand $d_i$. Our algorithm generates an unsplittable routing. In case the demands are arbitrary, it might not be possible to compare our algorithm to the best splittable routing. However, it is still possible to compare the performance to the best unsplittable routing, or to an optimal splittable routing that only uses paths with minimal
capacity of at least $d_i$.''
\end{comment}

\begin{comment}
    \item The problem was also investigated in distributed setting: RohitRohit~\cite{DBLP:conf/podc/AwerbuchFK09}, Leighton and Awerbuch~\cite{awerbuch1994improved, awerbuch1993simple}
\end{comment}

\subsection{Approaches for Online \MCF\ with High Demands}
We briefly discuss the weaknesses of approaches for solving the \ONMCF\
problem that rely directly on previous algorithms.

The algorithms in~\cite{AAP,BN06} route each request along a single
path. They require that the demand is smaller than the capacities.  In
order to apply these methods one should augment the capacities in advance so that the requested demand is bounded by the bottleneck along each path from the source to the destination.
This augmentation might be polynomial compared to the logarithmic augmentation requirement by our algorithm.

Another option is to split the requests into subrequests of small
demand so that each subdemand is smaller than the minimum capacity.
After that, a single-path online routing algorithm~\cite{AAP,BN06} can be
used to route each of these subrequests.  In this case, some of the
subrequests might be rejected, hence violating our all-or-nothing
requirement.

We define the \emph{granularity} of a flow as the smallest positive
flow along an edge in the network.  Let $\eps$ denote the granularity
of a flow.  One can formulate the multi-commodity problem as a packing
linear problem and apply the methods in~\cite{BN09,BNsurvey}.  The
edge capacity augmentation of these algorithms depends on the
$\log(\frac{1}{\eps})$, which might be unbounded.  For example,
consider the following network: (1)~The set of nodes is $V=\{u,v\}$,
(2)~there are two unit capacity parallel edges $(u,v)$. Consider a
request with demand $d_j=1+\eps$, for $\eps <1$. If the flow oracle
computes an all-or-nothing flow that routes flow of size $1$ on one
edge and $\eps$ on the other, then the granularity is $\eps$.

In order to solve this granularity problem, one can
apply~\cite{BN09,BNsurvey} and apply randomized rounding to obtain an
all-or-nothing solution with unit granularity.  Even in the
unit-demand case, this technique increases the competitive
%\footnote{Let $\opt$ denote the optimal value of a certain maximization problem. Let \alg\ denote an online algorithm for the same problem. Then, we say that  \alg\ is \emph{$\alpha$-competitive} if for all input sequences $\sigma$, $\alg(\sigma) \geq \frac{1}{\alpha} \cdot \opt(\sigma)$.}
ratio from $O(1)$ to $O(\log n)$ while the edge capacity augmentation
is $O(\log n)$.  Our result shows that an $O(1)$-competitive ratio is
achievable.

\begin{comment}
  Last, one can try apply~\cite{BN09,BNsurvey} to obtain an
  all-or-nothing algorithm, while \emph{over} satisfying flow requests
  and avoiding the ``small'' flow effect.  Since a customer will not
  pay more than he the aforementioned benefit, then the objective
  function of the packing formulation becomes non-linear - the primal
  dual-scheme does not apply in these cases.  We fill the addressed
  gap in this paper as stated in Section~\ref{sec:main}.
\end{comment}

\subsection{Techniques}
Our algorithm is based on the \emph{online primal-dual scheme}.  The
online primal-dual scheme by Buchbinder and
Naor~\cite{BNsurvey,BN09,BN06} invokes an ``min-weight'' path oracle.
The oracles considered in~\cite{BNsurvey,BN09,BN06} are either exact
oracles or approximate oracles.  Bansal et.
al~\cite{DBLP:conf/podc/BansalLNZ11} use bi-criteria oracles. Namely,
the oracles they considered are approximated and augment the edge
capacities.  We need tri-criteria oracles.

We extend the online primal-dual scheme so it supports
\emph{tri-criteria} oracles.  In the context of \MCF, the oracles
compute min-cost flow.  The three criteria of the these oracles are:
(1)~the approximation ratio, (2)~the capacity augmentation of the
edges, and (3)~the granularity of the computed flow.

Multiple criteria oracles were studied by Kolliopoulos and
Young~\cite{kolliopoulos2005approximation}. They presented bi-criteria
approximation algorithms for covering and packing integer programs.
Their algorithm finds an approximate solution while violating the
packing constraints. The granularity property is used
in~\cite{kolliopoulos2005approximation} to mitigate this violation.

\section{Problem Definition}\label{sec:problem}
Online multi-commodity flow (\ONMCF) is defined as follows.

\paragraph{The Network.} Let $G = (V,E)$ denote a directed graph,
where $V$ is the set of nodes and $E$ is the set of directed edges of
the network.  Let $n \eqdf |V|$, and $\res \eqdf |E|$.  Each edge $e
\in E$ has a capacity $c_e \geq 1$.

\paragraph{The Input.}
The online input is a sequence of requests $\sigma$, i.e., $\sigma =
\{r_j\}_{j\in \mathbb{N}^+}$.  Each flow request is a $4$-tuple $r_j =
(s_j,t_j,d_j,b_j)$.  Let $s_j,t_j \in V$ denote, respectively, the
source node and the target node of the $j$th request.  Let $d_j \geq
1$ denote the flow demand for the $j$th request.  Let $b_j \geq 1$
denote the benefit for the $j$th request.  We consider an online
setting, namely, the requests arrive one-by-one, and no information is
known about a request $r_j$ before its arrival.

\paragraph{The Output.}
The output is a multi-commodity flow $F = (f_{1},f_{2},\ldots)$.
For each request $r_j$, $f_{j}$ is a flow from $s_j$ to $t_j$.

\paragraph{Terminology.}
Let $|f_{j}|$ denote the amount of flow of $f_{j}$. Let $f_{j}(e)$
denote the $j$th flow along the edge $e \in E$.  Finally, for every $e
\in E$, $F^{(j)}(e) \eqdf \sum_{k=1}^j f_{k}(e)$, that is, the
accumulated flow along an edge $e$ after request $r_j$ is processed.
We say that an \MCF\ $F=(f_1,f_2,\ldots)$ \emph{fully serves} a
request $r_j$ if $|f_j|=d_j$.  We say that an \MCF\ $F$ \emph{rejects}
a request $r_j$ if $|f_j|=0$.  We say that an \MCF\ is
\emph{all-or-nothing} if each request is either fully served or
rejected.  An all-or-nothing \MCF\ is credited $b_j$ for each fully
served request $r_j$.
We say that an online \MCF\ (\ONMCF) algorithm
is \emph{monotone} if flow is never retracted.
We say that an online \MCF\ (\ONMCF) algorithm
is \emph{preemptive} if the flow $f_j$ of a fully served request $r_j$ is retracted entirely, i.e., $|f_j|=0$.
A monotone \ONMCF\
algorithm is, in particular, \emph{non-preemptive}.

\paragraph{The Objective.}
The goal is to compute an all-or-nothing \ONMCF\ that maximizes the total benefit of the served requests.

\subsection{The Main Result}\label{sec:main}
  We present an online algorithm for the \ONMCF\ problem that satisfies the following properties:
      \begin{enumerate}
        \item The algorithm is centralized and  deterministic.
        \item There is no limitation on demands. In particular, $\min_jd_j$ may exceed $\max_e c_e$.
        \item The algorithm is all-or-nothing.
        \item The online algorithm \alg\ competes with an all-or-nothing offline optimal algorithm.
        \item The algorithm is $(1+\delta)$-competitive, for a constant $\delta \in (0,1]$.
        \item The algorithm violates the capacity constraints by an $O(\log n)$ factor.
        \item The algorithm is non-preemptive and monotone.
      \end{enumerate}

\noindent
For a vector $x=(x_1,\ldots,x_k)$, let $x_{\min}\eqdf \min_i x_i$. Similarly,
$x_{\max}\eqdf \max_i x_i$.
The main result is formalized in Theorem~\ref{thm:main2}.

\begin{theorem}[\textbf{Main Result}]\label{thm:main2}
Let $\gamma$ denote a constant. Assume that:
 \begin{enumerate}
    \item $1\leq b_{\min} \leq b_{\max} \leq O(n^\gamma)$,
    \item $1\leq c_{\min} \leq c_{\max} \leq O(n^\gamma)$,
    \item $1\leq d_{\min}$.
 \end{enumerate}
 Then, Algorithm~\ref{alg:alg} is a non-preemptive, monotone, online
 algorithm for the \ONMCF\ problem that computes an all-or-nothing
 multi-commodity flow that is $(O(1),O(\log n))$-competitive.
\end{theorem}

\section{Online Packing and Covering Formulation}
In this section we present a sequence of packing linear programs (LPs)
that correspond to the \ONMCF\ problem. We also present covering
linear programs. We refer to the covering programs as the primal LPs and to the packing programs as the dual LPs.

\subsection{Flow Polytopes}\label{sec:A}
We define polytopes of flows that correspond to the requests $\{r_j\}_{j \in \mathbb{N}^+}$ as follows.
\begin{definition}\label{def:polytope}
  For every $r_k = (s_k,t_k, d_k,b_k)$, let $\Pi_k(\mu)$ denote the polytope of \emph{unit} flows $f$ from $s_k$ to $t_k$ in $G$ that satisfy: $\forall e \in E: f(e) \leq \mu \cdot \frac {c_e}{d_k}$.
\end{definition}
\noindent
We refer to $\Pi_k(1)$ simply by $\Pi_k$.
Let $V(\Pi_k(\mu))$ denote the set of extreme points of $\Pi_k(\mu)$.
\begin{definition}
We say that request $r_k$ is \emph{$\mu$-feasible} if  $\Pi_k(\mu) \neq \emptyset$.
We say that request $r_k$ is \emph{feasible} if  $\Pi_k \neq \emptyset$.
\end{definition}
\noindent
Note that a request $r_j$ is $\mu$-feasible if and only if the
capacity of the minimum cut that separates $s_k$ from $t_k$ is at
least $\frac{d_k}{\mu}$.  In particular, a request $r_j$ may be feasible
even if $d_j > \max_e c_e$.
\subsection{Packing and Covering Formulation}
For every prefix of requests $\{r_k\}_{k=1}^j$ we define a primal linear program $\plp(j)$ and a dual linear program $\dlp(j)$.
The LP's appear in Figure~\ref{fig:LPframe}.

\begin{figure}
\centering
\begin{tabular}{| l |}
\hline
    $\underline{\plp(j)}:$\\
    \begin{minipage}{0.67\textwidth}
     \begin{center}
        \begin{eqnarray*}
        \min~  \sum_{k=1}^j d_k \cdot z_k +\sum_{e\in E} c_e \cdot x_e &\text{s.t.}& \\
         \forall k\in [1,j] ~\forall f \in V(\Pi_k): z_k + \sum_{e\in E}x_e \cdot f(e) &\geq& \frac{b_k}{d_k}\\
         x,z &\geq& \vec{0}
        \end{eqnarray*} \\
        (I)
     \end{center}
    \end{minipage}
\\
\hline
    $\underline{\dlp(j)}:$\\
    \begin{minipage}{0.67\textwidth}
     \begin{center}
        \begin{eqnarray*}
         \max~ \sum_{k=1}^j \sum_{f \in V(\Pi_k)} \frac{b_k}{d_k}  \cdot y_f &\text{s.t.}&\\
         \forall e\in E: \sum_{k=1}^j\sum_{f\in V(\Pi_k)}f(e) \cdot y_f&\leq& c_e\text{ \footnotesize (Capacity Constraints.)} \\
         \forall k\in [1,j]: \sum_{f \in V(\Pi_k)}y_f &\leq & d_k\text{ \footnotesize (Demand Constraints.)}\\ y &\geq& \vec{0}
        \end{eqnarray*}
        \\(II)
     \end{center}
    \end{minipage}
    \\
\hline
\end{tabular}
\caption{
(I) The primal LP $\plp(j)$.
(II) The dual LP $\dlp(j)$.}
   \label{fig:LPframe}
\end{figure}

The packing program $\dlp(j)$ has a variable $y_f$ for every flow $f
\in \bigcup_k V(\Pi_k)$ and two types of constraints: \emph{demand
  constraints} and \emph{capacity constraints}.  The capacity
constraints require that the load on every edge $e$ is at most $c_e$.
The demand constraints require that the conical combination of unit
flows in $V(\Pi_k)$ is a flow of size at most $d_k$.

The covering program $\plp(j)$ has a variable $x_e$ for every edge $e
\in E$, and a variable $z_k$ for every request $r_k$, where $k \leq
j$.  It is useful to view $x_e$ as the cost of a unit flow along $e$.

\section{The Online Algorithm \alg}\label{sec:gen}
In this section we present the online algorithm \alg.

\subsection{Preliminaries}
The algorithm maintains the following variables: (1)~For every edge
$e$ the primal variable $x_e$, (2)~for every request $r_j$ the primal
variable $z_j$, and (3)~the multi-commodity flow $F$.  The primal
variables $x,z$ are initialized to zero.  The \MCF\ $F$ is initialized
to zero as well.

\paragraph{Notation.}
Let $x^{(j)}_e$ denote the value of the primal variable $x_e$ after request $r_j$ is processed by \alg.

For every request $r_j$, let $\cost_j(f)$ denote the $x$-cost of a flow $f$,  formally:
% \in \Pi_j(\mu)$
$$\cost_j(f)\eqdf \sum_e x_e^{(j-1)}\cdot f(e)\:.$$

For every flow $f$, let $w(f)$ denote the sum of the flows along the edges,  formally:
$$w(f) \eqdf \sum_e f(e)\:.$$

Let $F^{(k)}$ denote the \MCF\ $F$ after request $r_k$ is processed.
Let $\val_j(F)$ denote the benefit of \MCF\ $F$ after request $r_j$ is processed, formally:
\[
    \val_{j}(F)\eqdf \sum\{b_i \mid i \leq j,r_i \text{ is fully served by }F^{(j)}\}\:.
\]

Let $\valp_j(x,z)$ denote the objective function's value of $\plp(j)$ for a given $x$ and $z$, formally:
\[
    \valp_j(x,z) \eqdf \sum_{k=1}^j d_k \cdot z_k +\sum_{e\in E} c_e \cdot x_e^{(j)}\:.
\]
Let $\I$ denote an all-or-nothing offline optimal \MCF\ w.r.t input sequence $\sigma=\{r_j\}_j$.
%Analogously, let $\val_j(\I)$ denote the benefit of flow $\I$ after $r_j$ is processed.

\begin{definition}
  An \MCF\ $F=(f_1,f_2,\ldots)$ is \emph{$(\alpha,\beta)$-competitive}
  with respect to a sequence $\{r_j\}_j$ of requests if for every $j$:
  \begin{enumerate}[(i)]
  \item $F$ is $\alpha$-competitive: $\val_j(F) \geq \frac 1{\alpha} \cdot
    \val_j(\I)$.
  \item $F$ is $\beta$-feasible: for every $e \in E$, $F^{(j)}(e) \leq \beta \cdot c_e$.
  \end{enumerate}
\end{definition}

\begin{definition}
  An \MCF\ $F=(f_1,f_2,\ldots)$ is \emph{all-or-nothing} if each
  request $r_j$ is either fully served by $F$ or it is rejected by $F$
  (i.e., $|f_j|\in \{0,d_j\}$).
\end{definition}

\subsection{Description}
Upon arrival of a request $r_j$, if the request is not feasible, then the algorithm rejects it upfront.
Otherwise, if the request is feasible, then \alg\ invokes a  tri-criteria oracle. The oracle returns a unit-flow $f_j$ for $r_j$.

If the cost of the oracles's flow is ``small enough'', then the request is accepted as follows: (1)~the flow $F$ is updated by adding the oracle's unit-flow $f_j$ times the required demand $d_j$, (2)~the primal variables $x_e$, for every edge $e$ that the flow $f_j$ traverses, are  updated.

If the flow is ``too expensive'', then the request is rejected and no updates are made to the primal variables and to the \MCF\ $F$.

The listing of the online algorithm \alg\ appears in Algorithm~\ref{alg:alg}.

\newcounter{saveenum}
\begin{algorithm}
    \textbf{Initialize:} $z \leftarrow 0, x \leftarrow 0, F \leftarrow 0$.\\
    \textbf{Upon arrival} of request $r_j = (s_j,t_j,d_j,b_j)$, for $j\geq 1$:
        \begin{enumerate}[1)]
        \item If $r_j$ is not feasible (i.e., $\Pi_j =\emptyset$), then
          \textbf{reject} $r_j$ and skip the remaining lines.
            \item \label{step:general_oracle} \label{step:oracle}
                    $f_j \leftarrow \oracle(x,r_j )$~~~~~\{The oracle is a  $(\lambda,\mu,\eps)$-criteria.\}
            \item \label{step:general_th} \label{step:2}
                    If $d_j \cdot \cost_j(f_j) < \lambda \cdot b_j$,
                    \setcounter{saveenum}{\value{enumi}}
                    \begin{enumerate}[1)]
                    \setcounter{enumii}{\value{saveenum}}
                    \item then \textbf{accept} $r_j$
                    \setcounter{saveenum}{\value{enumii}}
                      \begin{enumerate}[1)]
                      \setcounter{enumiii}{\value{saveenum}}
                      \item \label{step:2a} $F \leftarrow F+d_j \cdot
                        f_j$~~~~~\{Updating the multi-commodity flow.\}
                      \item $z_j \leftarrow \frac{b_j}{d_j} -
                        \frac{\cost_j(f_j)}{\M}$ \label{step:2c}
                      \item \label{step:general_xe_update}
                        \label{step:2b} $\forall e: f_j(e) > 0$:
                        \begin{align*}
                          L_j(e) &\eqdf \frac{d_j \cdot f_j(e)}{\M \cdot c_e}  \\
                          x_{e} &\gets x_{e}\cdot
                          2^{L_j(e)}+\frac{1}{d_j\cdot w(f_j)}\cdot
                          \left(2^{L_j(e)}-1\right)
                        \end{align*}
                      \end{enumerate}
                      \setcounter{saveenum}{\value{enumiii}}
                    \end{enumerate}
                    \setcounter{enumi}{\value{saveenum}}
                 \item Else \textbf{reject} $r_j$
            \end{enumerate}
\caption{\alg: Online multi-commodity flow algorithm. The algorithm receives a sequence of requests and outputs a multi-commodity flow $F$.}\label{alg:alg}
\end{algorithm}

\subsection{The Oracle}
The oracle description is as follows:
\begin{enumerate}[(i)]
  \item \textbf{Input:} Request $r_j$, edge capacities $\frac{c_e}{d_j}$, and edge costs $x^{(j-1)}:E \rightarrow \mathbb{R}^{\geq0}$.
  \item \textbf{Output:} A unit-flow $f$ from $s_j$ to $t_j$.
\end{enumerate}

\medskip \noindent
Let $\opt_j$ denote the min-cost flow in $\Pi_j$ w.r.t. the costs $x_e$, formally: $$\opt_j \eqdf  \arg\min\{\cost_j(f) : f \in \Pi_j \}\:.$$
Note that: (1)~$\opt_j$ is well defined because $\Pi_j \neq \emptyset$, and (2)~the edge capacities in $\Pi_j$ are $\frac{c_e}{d_j}$.

The oracles in our context are tri-criteria, as formalized in the following definition.
\begin{definition}[\textbf{Oracle Criteria}]
We say that an oracle is \emph{$(\lambda,\mu,\eps)$-criteria}, if the oracle outputs a flow $f$ that satisfies the following properties:
  \begin{enumerate}[(i)]
  \item \textbf{($\lambda$-Approximation.)} $\cost_j(f) \leq \lambda \cdot
    \cost_j(\opt_j)$. %where $\lambda \geq 1$,
  \item \textbf{($\mu$-Augmentation.)} $f\in \Pi_j(\mu)$.
  \item \textbf{($\eps$-Granular.)} $f(e)>0 \Rightarrow f(e) \geq \eps$.
  \end{enumerate}
\end{definition}

\subsubsection{A Tri-criteria Oracle for Minimum Cost Flow}\label{sec:oracle}
The oracle's listing is as follows.
\paragraph{The Oracle Outline.}
\begin{enumerate}
  \item Let $f \leftarrow \opt_j$.
  \item Decompose $f$ to at most $m$ flow paths $\{f_{1},\ldots, f_{m}\}$.
  \item Remove each flow path $f_{\ell}$ such that $|f_{\ell}| < \frac{1}{{2\res}^2}$.
  \item Let $g$ denote the removed flow from $f$.
  \item Scale every remaining flow path $f_\ell$ (i.e., $|f_\ell| \geq \frac{1}{2\res^2}$) as follows:
      \[
        f_{\ell} \leftarrow f_{\ell} \cdot \left( 1 + \frac{|g|}{|f|-|g|} \right)
      \]
\end{enumerate}
The proof of the following lemma appears in Appendix~\ref{sec:oracleproof}.
\begin{lemma}\label{lemma:minfloworacle}
  The oracle is $(2,2,\frac{1}{{2\res}^2})$-criteria algorithm.
\end{lemma}
Lemma~\ref{lemma:minfloworacle} justifies using the following parameters: $\lambda = \mu = 2$, and $\eps = \frac{1}{2\res^2}$.

\section{Analysis}~\label{sec:analysis}
The following observation is proved by the fact that $\frac 1z \cdot(2^{z}-1)$ is monotone increasing for $z >0$.
\begin{observation}\label{obs:ineq}
Let $c \in \mathbb{R}^{>0}$, then $$\forall x\in [0,c]: c \cdot(2^{x/c}-1) \leq x \:.$$
\end{observation}
\begin{comment}
  \begin{proof}
  It suffices to prove that $f(z) = \frac 1z \cdot(2^{z}-1)$ is monotone increasing for $z\in (0,\mu]$.
  Indeed, the first derivative of $f$ is positive for all $z \in \mathbb{R}^{>0}$.
  Since $f$ is monotone increasing it follows that,
  $$\forall z\in [0,\mu]: \frac 1z \cdot(2^{z}-1) \leq \frac{1}{\mu}\cdot (2^\mu-1) \:,$$ and the observation follows.
%  Proof that f'(z) is positive for z>0:
%    For $z \geq 1/ \ln 2$ the 1st derivative is $\geq 0$ for $1/\ln 2 > z > 0$ it follows that $f'(z)\cdot z^2 > z\cdot 2^z - e+1 > z \cdot 2^z > 0$
\end{proof}
\end{comment}
\begin{observation}\label{obs:2}
If $L_j(e) \leq 1$, then
    \[
\left(2^{L_j(e)}-1\right)\cdot c_e \leq \frac{1}{\M}\cdot d_j \cdot f_j(e).
\]
  \end{observation}
  \begin{proof}
By Observation~\ref{obs:ineq} and since $L_j(e) \leq 1$, it follows that
\[
\left(2^{L_j(e)}-1\right)\cdot \M \cdot c_e \leq d_j \cdot f_j(e),
\]
and the observation follows.
  \end{proof}
\paragraph{Notation.}
%Let $\cmax \eqdf \max_e c_e$, and $ \bmax \eqdf \max_j b_j$, and $ \dmax \eqdf \max_j d_j$.
Let
\begin{eqnarray*}
\alpha & \eqdf & 1 + \frac{1}{\M} \leq 2\:,\\
\beta &\eqdf & \M\cdot\log_2\left(1+{\res}^2\cdot
    \frac {3\cdot \lambda \cdot \cmax\cdot \bmax}{\eps} \right)\:.
\end{eqnarray*}

In the following theorem we prove that \alg\ is an all-or-nothing $(\alpha,\beta)$-competitive, non-preemptive and monotone online algorithm.
\begin{theorem}~\label{thm:main result}
Assume that:
%\begin{inparaenum}[(i)]
\begin{enumerate}
\item $b_{\min}, c_{\min}, d_{\min} \geq 1$.
\item The oracle is $(\lambda,\mu,\eps)$-criteria.
\end{enumerate}
%\end{inparaenum}
Then \alg\ is non-preemptive, monotone, online algorithm for the \ONMCF\ problem that computes
an all-or-nothing multi-commodity flow that is
$(\alpha,\beta)$-competitive.
\end{theorem}
\begin{proof}
  The \alg\ algorithm rejects upfront requests that are not
  feasible. These requests are also rejected by $\I$, hence it
  suffices to prove $(\alpha,\beta)$-competitiveness w.r.t fractional
  offline optimal algorithm over the feasible requests.  We now
  prove $\alpha$-competitiveness and $\beta$-feasibility.

\paragraph{$\alpha$-competitiveness.}
First, we prove $\alpha$-competitiveness.  Let $\PR \eqdf \valp_j(x,z)
- \valp_{j-1}(x,z)$, and $\DU \eqdf \val_{j}(F)-\val_{j-1}(F)$.  We
begin by proving that $\PR \leq \alpha\cdot \DU$ for every request
$r_j$.

Recall that $x^{(j)}_e$ denotes the value of the primal variable $x_e$
after $r_j$ is processed.  If $r_j$ is rejected then $\PR = \DU =0$
and the claim holds.  If $r_j$ is accepted, then $\DU = b_j$ and $\PR
= \sum_e (x^{(j)}_{e}-x_e^{(j-1)})\cdot c_e +d_j\cdot z_{j}$. Let
$f_j$ denote the output of the oracle when dealing with request $r_j$,
i.e., $f_j \leftarrow \oracle(x^{(j-1)},r_j)$. Indeed,
    \begin{eqnarray}
    \label{eq:deltax_e}
 \sum_{e}\left(x^{(j)}_{e}-x_e^{(j-1)}\right)\cdot c_e
         & = & \sum_{e} \left[x^{(j-1)}_{e}
          \cdot\left(2^{L_j(e)}-1\right)+
          \frac{1}{d_j\cdot w(f_j)}\cdot \left(2^{L_j(e)}-1\right)\right]\cdot c_e \nonumber\\
         & = &  \sum_{e} \left(x^{(j-1)}_{e}+\frac{1}{d_j\cdot w(f_j)}\right) \cdot
            \left(2^{L_j(e)}-1\right)\cdot c_e  \nonumber\\
         & \leq & \sum_{e} \left(x^{(j-1)}_{e}+\frac{1}{d_j\cdot w(f_j)}\right) \cdot
             \frac{d_j\cdot f_j(e)}{\M}\nonumber\\
         & =& \frac{d_j \cdot \cost_j(f_j)}{\M}+\frac{1}{\M}
, \label{eq:delta x}
    \end{eqnarray}
where the third inequality holds since the oracle is $\mu$-augmented and by Observation~\ref{obs:2}. Hence, Equation~\ref{eq:delta x} and Step~6 of \alg\ imply that:%\ref{step:2c}
\begin{eqnarray*}
    \PR & \leq &  \frac{d_j \cdot \cost_j(f_j)}{\M}+\frac{1}{\M}+d_j\cdot z_j \nonumber\\
    & =&  \frac{d_j \cdot \cost_j(f_j)}{\M}+\frac{1}{\M}+d_j\cdot \left(\frac{b_j}{d_j} - \frac{\cost_j(f_j)}{\M}\right)\\ %\label{eq:delta p}
    & = & \frac{1}{\M} + b_j \leq \alpha \cdot b_j\:,
 \end{eqnarray*}
where the last inequality holds since $b_j \geq 1$.

 Since $\DU = b_j$ it follows that
 \begin{align}
 \PR \leq \alpha \cdot \DU\:, \label{eq:comp}
 \end{align}
 as required.

Initially, the primal variables and the flow $F$ equal zero. Hence, Equation~\ref{eq:comp} implies that:
\begin{eqnarray}
   \valp_j(x,z) &\leq&  \alpha\cdot \val_j(F)\:. \label{eq:comp2}
\end{eqnarray}

We now prove that, the primal variables $\{x^{(j)}_{e}\}_{e} \cup \{z_{i}\}_{i\leq j}$ constitute a feasible solution for $\plp(j)$:
\begin{enumerate}
  \item  If $r_j$ is rejected, then $\cost_j(f_j) \geq \lambda\cdot \frac{b_j}{d_j}$. Since the oracle is $\lambda$-approximate it follows that for every $f' \in V(\Pi_j)$:
    \[
        \cost_j(f') \geq \cost_j(\opt_j) \geq \cost_j(f_j)/\lambda \geq \frac{b_j}{d_j}\:.
    \]
It follows that the primal constraints are satisfied in this case.
  \item If $r_j$ is accepted, then $\cost_j(f_j) < \lambda\cdot\frac{b_j}{d_j}$.
    Since $z_j=\frac{b_j}{d_j} - \frac{\cost_j(f_j)}{\M}$ it follows that for every $f' \in V(\Pi_j)$:
    \begin{eqnarray*}
      z_j+\cost_j(f') & \geq & \frac{b_j}{d_j} - \frac{\cost_j(f_j)}{\M} +\frac{\cost_j(f_j)}{\lambda}
       \geq  \frac{b_j}{d_j}\:.
    \end{eqnarray*}
    We conclude that the primal constraints are satisfied in this case
    as well.
\end{enumerate}

The first $j$ flows of the optimal offline multi-commodity flow $F^*$
are clearly a feasible solution to $\dlp(j)$.  The value of this
solution equals $\val_{j}(F^*)$.  Since the primal variables
constitute a feasible primal solution, weak duality implies that:
\begin{align*}
  \val_{j}(F^*) &\leq \valp_j(x,z)\:.
\end{align*}
Hence, by Equation~\ref{eq:comp2}, it follows that:
\[
    \val_j(F) \geq  \frac{1}{\alpha} \cdot \val_{j}(F^*),
\]
which proves that \alg\ is $\alpha$-competitive.  \medskip

\paragraph{$\beta$-feasibility.}
We now prove $\beta$-feasibility, i.e., for every $r_i$ and for every $e
\in E$, $F^{(i)}(e) \leq \beta \cdot c_e$.

We prove a lower bound and an upper bound on $x_e$ in the next two lemmas.
Let $r_j$ denote the index of the last request.
Let
\[
W \eqdf \max\{d_k \cdot w(f_{k})  : 0\leq k \leq j\}.
\]
\begin{lemma} \label{lemma:induct_xe}
  For every edge $e$,
    \begin{align*}
    x_e \geq  \frac{1}{W} \cdot \left(2^{F(e)/\M \cdot c_e}-1\right)\:.
    \end{align*}
\end{lemma}

\begin{proof}
Recall that  $x^{(k)}$ (resp. $F^{(k)}$)
  denote the value of $x$ (resp. $F$) after request $r_k$ is
  processed.
We prove by induction on $k\leq j$ that
\begin{align}
  x^{(k)}_e \geq \frac{1}{W} \cdot \left(2^{F^{(k)}(e)/\M \cdot
    c_e}-1\right). \label{eq:x lower bound}
\end{align}
The induction basis, for $k=0$, holds because both sides equal zero.

\medskip\noindent \emph{Induction step}: Note that if $r_k$ is
rejected, then both sides of Equation~\ref{eq:x lower bound} remain
unchanged, and hence Equation~\ref{eq:x lower bound} holds by the
induction hypothesis.  We now consider the case that $r_k$ is
accepted.

The update rule in Step~7 of \alg\ implies that%\ref{step:2b}

    \begin{eqnarray*}
      x^{(k)}_{e} & = &
      x^{(k-1)}_{e}\cdot 2^{\frac{d_k\cdot f_k(e)}{\M \cdot c_e }}
     +\frac{1}{d_j\cdot w(f_k)}\cdot
    \left(2^{\frac{d_k\cdot f_k(e)}{\M \cdot c_e} }-1\right)
\\
              & \geq &  \frac{1}{W} \cdot \left(2^{\frac{F^{(k-1)}(e)}{\M \cdot
    c_e}}-1\right) \cdot 2^{\frac{d_k\cdot f_k(e)}{\M \cdot c_e}}
     +\frac{1}{d_j\cdot w(f_k)}\cdot
    \left(2^{\frac{d_k\cdot f_k(e)}{\M \cdot c_e}}-1\right)\\
&\geq &
\frac{1}{W} \cdot \left(2^{\frac{F^{(k)}(e)}{\M \cdot
    c_e}}-1\right).
    \end{eqnarray*}
    The lemma follows.
\end{proof}

    \begin{lemma}\label{lemma:xe ub}
      For every accepted request $r_k$, if $f_k(e)>0$, then
      $$x^{(k)}_e  \leq \frac{3 \cdot \lambda \cdot b_k} {\eps \cdot d_k}\:.$$
    \end{lemma}
    \begin{proof}
      Since $r_k$ is accepted, we have $d_k\cdot \cost_k(f_k) <
      \lambda \cdot b_k$.  By $\eps$-granularity of the oracle,
      $\cost_k(f_k) \geq x_e^{(k-1)}\cdot \eps$. It follows that
      $x^{(k-1)}_e \leq \frac{\lambda \cdot b_k}{d_k \cdot \eps}$. By the update rule for $x_e$, we have:
    \begin{align*}
    x^{(k)}_e \leq \frac{\lambda \cdot b_k}{d_k \cdot \eps} \cdot 2^{L_k(e)}+\frac{1}{d_k \cdot w(f_k)}\cdot \left(2^{L_k(e)}-1\right)\:.
    \end{align*}
    Since the oracle is $\mu$-augmented, $L_k(e) \leq 1$.
    In addition, since the oracle is $\eps$-granular, $w(f_j)\geq \eps$.
    \begin{eqnarray*}
      x^{(k)}_e &\leq & \frac{\lambda \cdot b_k}{d_k \cdot \eps} \cdot 2 + \frac {1}{d_k \cdot \eps} \cdot (2-1) \nonumber\\
        &\leq &\frac{3 \cdot \lambda \cdot b_k} {d_k \cdot \eps}\:. %\label{eq:bound x}
    \end{eqnarray*}
    \end{proof}

Lemma~\ref{lemma:induct_xe} and Lemma~\ref{lemma:xe ub} imply that:
\[
\frac{1}{W} \cdot \left(2^{F^{(k)}(e)/\M \cdot c_e}-1\right)
\leq
\max_{k\leq j}
\frac{3 \cdot \lambda \cdot b_k} {\eps \cdot d_k}\:.
\]

Hence,
\begin{align}
  F^{(k)}(e) \leq c_e \cdot \M \cdot \log_2\left(1+W \cdot \max_{k}
\frac{3 \cdot \lambda \cdot b_k} {\eps \cdot d_k}\right)\:. \label{eq:aug}
\end{align}

Since
\begin{inparaenum}[(i)]
\item
$W \leq \res \cdot \dmax$,
\item $\dmax \leq \res \cdot \cmax$, and
\item $d_{\min} \geq 1$,
\end{inparaenum}
it follows that,
\begin{align*}
  F^{(k)}(e) &\leq \beta \cdot c_e\:,
\end{align*}
for every $k$, as required.
\end{proof}

This concludes the proof of Theorem~\ref{thm:main result}.
Theorem~\ref{thm:main2} follows directly from Theorem~\ref{thm:main
  result} and Lemma~\ref{lemma:minfloworacle}.

\begin{rem}
Let $\bpb_k$ denote the benefit-per-bit of request $r_k$, i.e., $\bpb_k \eqdf \frac{b_k}{d_k}$. Let $\bpb_{\max} \eqdf \max_k\bpb_k$.
Instead of $\beta$, the augmentation can be also bounded by:
$$\M \cdot\log_2\left(1+W \cdot \frac{3 \cdot \lambda }{\eps} \cdot \bpb_{\max}\right)\:.$$
\end{rem}

\section{Mixed Demands}
One may consider a mixed case of low and high demands. A flow request
with high demand has to be split into multiple paths.  Splitting a
stream of packets along multiple paths should avoided, if possible,
because it complicates implementation in nodes where flow is split,
may cause packets to arrive out-of-order, etc. Thus, one may require
not to split requests with low demand. Formally, a request has low
demand if $d_j \leq \cmin$; otherwise, it has a high demand.

An online algorithm for mixed demands can be obtained by employing two
oracles: (1)~A tri-criteria oracle for the high demands.  This oracle
may serve a flow request by multiple paths. (2)~An exact (shortest
path) oracle for low demands. This oracle must serve a flow request
by a single path.

\begin{theorem}
  There exists a non-preemptive, monotone, online algorithm for the
  \ONMCF\ problem with mixed demands that computes an all-or-nothing
  multicommodity flow that is $(O(1),O(\log n))$-competitive.
\end{theorem}

\begin{proof}[Proof sketch]
  The proof is based on the feasibility of the primal LP and on the
  bounded gap between $\DU$ and $\PR$. These two invariants are
  maintained regardless of the oracle that is invoked. The proof for
  the case of small demands appears in~\cite{BNsurvey}. The augmentation
  of the capacities are determined by the oracle with the ``worst''
  parameters. Because the exact oracle is $(1,1,1)$-criteria, it is
  also $(\lambda,\mu,\epsilon)$-criteria. Thus, the  augmentation factor
  $\beta$ is determined by the approximate oracle.
\end{proof}
\section{Further Extensions}
\paragraph{Requests with known durations.}
The algorithm can be extended to deal with flow requests with known
durations.  For the sake of simplicity, the flow requests in this
paper are permanent, namely, after arrival, a request stays forever.
Using previous techniques~\cite{AAP,BN06,DBLP:conf/icdcn/EvenMSS12}, our algorithm can be adapted to deal also
with the important variant of known durations. In this variant, each
request, upon arrival, also has an end-time.  The competitive ratio
for known durations when the requests are a logarithmic fraction of
the capacities is $O(\log (nT))$, where $T$ denotes the longest
duration~\cite{AAP}. In fact, the primal-dual method in~\cite{BN06}
can be extended to the case of routing requests with known durations
(see~\cite{DBLP:conf/icdcn/EvenMSS12}). Thus, for known durations, if
the demands are bounded by the minimum capacity, then the primal-dual
method yields an online algorithm, the competitive ratio of which is
$(O(1),O(\log (nT)))$. One can apply a tri-criteria oracle with
granularity $O(n^{-2})$, to obtain an $(O(1),O(\log
(nT)))$-competitive ratio for known durations even with high demands.

%%%%
\paragraph{All-or-nothing machine scheduling.}
A simple application of our algorithm is the case of maximizing
throughput in an online job all-or-nothing scheduling problem on
unrelated machines.  The variant in which the objective is to minimize
the load was studied by Aspnes et al.~\cite{aspnes1997line}. We, on
the other hand, focus on maximizing the throughput.

Jobs arrive online, and may be assigned to multiple machines
immediately upon arrival. Moreover, a job may require specific subset
of machines, i.e., restricted assignment.  The increase in the load of
a machine when a job is assigned to it is a function of the machine
and the fraction of the job that is assigned to it.  Formally, Let
$\tau_j(e) \in [0,1]$ denote the ``speed up'' of machine $e$ when
processing job $j$, that is, one unit of job $e$ on machine $j$ incurs
a an additional load of $\tau_j(e)$ on machine $e$.  The reduction is
to network of $m$ parallel edges, one edge per machine. The capacity of
each edge equals the capacity of the corresponding machine.

Large jobs need to be assigned to multiple machines, while small jobs
may be assigned to a single machine (as in~\cite{aspnes1997line}).
In this case our algorithm is
$(O(1),O(\frac{\log m}{\min_{j,e}\tau_j(e)}))$-competitive,
where $m$ is the number of machines.

% High Demands, Granularity, restricted, related, identical, unrelated, max -th

\begin{comment}
 The scheme in this paper can be applied in the context of machine load balancing. In this case each request is simply a job that needs to be done. Each job has a manufacturing demand of resources from the ``factory''. Since there is not a ``partially done'' job, the factory has to work in an all-or-nothing approach. The factory is modeled by the graph $G=(V,E)$, which consists of two vertices, and $m$ is the number of its machines. Similarly to edge capacities, each machine has its manufacturing capacity.
 We are interested in maximizing the benefit gained by served jobs.
Recall that the manufacturing demands of each job are high,  hence each job must be ``broken'' into sub-jobs (so this job could be manufactured by several machines in parallel). Now, breaking each job arbitrarily to sub-jobs is nonrealistic, hence defining the granularity of sub-jobs is essential, i.e., you cannot manufacture half a bolt.
\end{comment}

\bibliographystyle{alpha}
\bibliography{bibgran}
\appendix
\section{Proof of Lemma~\ref{lemma:minfloworacle}}\label{sec:oracleproof}
In this section we prove the following lemma.
\paragraph{Lemma~\ref{lemma:minfloworacle}}\emph{
  The oracle is $(2,2,\frac{1}{{2\res}^2})$-criteria algorithm.}
\begin{proof}
  Throughout this proof we refer to flow path that are not removed, simply by `flow paths'.

  First, we prove that the oracle outputs a unit flow.
  For every flow path $f_\ell$ such that $|f_\ell| \geq \frac{1}{2{\res}^2}$, let $f_\ell^{(s)}$ denote the scaled flow along it.
  The sum of the flows, along the scaled flow paths equals:
  \[
    \sum_\ell f^{(s)}_\ell = \sum_\ell f_\ell \cdot \left( 1+ \frac{|g|}{|f|-|g|} \right)=|f|-|g|+|g|=|f|\:.
  \]
  Hence, the oracle outputs a unit flow as required.

  The oracle is $\frac{1}{2{\res}^2}$-granular by construction.

  We prove that the oracle is $2$-augmented.
  Note that  $|g| \leq \frac{\res}{2{\res}^2}=\frac{1}{2{\res}}$.
  Hence,
  \[
  \frac{|g|}{|f|-|g|} \leq \frac{\frac{1}{2{\res}}}{1-\frac{1}{2{\res}}} = \frac{1}{2\res-1}\:.
  \]
  It follows that the flow along every edge is augmented by at most
  \[
    \left( 1+ \frac{|g|}{|f|-|g|} \right) \leq \left( 1+ \frac{1}{2\res-1} \right) < 2\:.
  \]
  Hence, $f \in \Pi_j(2)$, as required.

  Moreover, the flow along every scaled flow path $f^{(s)}_\ell$ satisfies for every $e \in E$:
  \[
    f^{(s)}_\ell(e) \leq f_\ell(e) \cdot \left( 1+ \frac{1}{2\res-1} \right)\:,
  \]
  Hence,
  \[
    \cost_j(f^{(s)}_\ell) \leq \cost_j(f_\ell)\cdot \left( 1+ \frac{1}{2\res-1} \right)\:,
  \]
  which proves $2$-approximation of the oracle.
\end{proof}

\end{document}